\documentstyle[12pt]{article}

\begin{document}

\centerline{\Large A Lagrangian for Electromagnetic}
\vspace{.4cm}
\centerline{\Large Solitary Waves in Vacuum}
\vspace{.9cm}
\centerline{\large Daniele Funaro}
\vspace{.3cm}
\centerline{\small Dipartimento di Matematica}
\vspace{.1cm}
\centerline{\small Universit\`a di Modena e Reggio Emilia}
\vspace{.1cm}
\centerline{\small Via Campi 213/B, 41125 Modena (Italy)}
\vspace{.1cm}
\centerline {\small E-mail: daniele.funaro@unimore.it}
\vspace{.8cm}

\begin{abstract}
A system of equations, describing the evolution of electromagnetic
fields, is introduced and discussed. The model is strictly related
to Maxwell's equations. As a matter of fact, the Lagrangian is the
same, but the variations are subjected to a suitable constraint.
This allows to enlarge the space of solutions, including for
example solitary waves with compact support. In this way, without
altering the physics, one is able to deal with vector waves as
they were massless particles. The main properties of the model,
together with numerous exact explicit solutions are presented.
\end{abstract}

{Keywords: Solitons, Electromagnetism, Wave-fronts, Lagrangian}

{PACS: 41.20.Jb, 42.15.-i, 02.30.Xx}

\setcounter{equation}{0}

\section{The modelling equations}

In \cite{funaro}, the theory of electromagnetism has been
reviewed by introducing a suitable nonlinear set of equations,
allowing for a very extended space of solutions, basically
including any type of {\sl solitary wave}. The new formulation
provides a far more accurate description of wave phenomena,
including their self-interactions. We reorganize here part of the
material, providing additional insight.
\par\smallskip

We are mainly concerned with the evolution of pure electromagnetic
waves in vacuum. Denoting by $c$ the speed of light, in Minkowski
space the model equations read as follows:
\begin{equation}\label{eq:vare}
{\partial {\bf E}\over \partial t}~=~ c^2 {\rm curl} {\bf B}~
-~({\rm div}{\bf E}) {\bf V}
\end{equation}
\begin{equation}\label{eq:varb}
{\partial {\bf B}\over \partial t}~=~ -{\rm curl} {\bf E}~
\end{equation}
\begin{equation}\label{eq:varv}
{\bf E}~+~{\bf V}\times{\bf B}~=~0
\end{equation}
  Here, ${\bf E}$ and ${\bf B}$ are
the usual electric and magnetic fields, while ${\bf V}$  is a new
velocity field. Note that there are 3 vector unknowns and 3 vector
equations. We will require ${\bf V}$ to be of constant norm. More
exactly, we impose:
\begin{equation}\label{eq:normv}
\Vert{\bf V}\Vert~=~c
\end{equation}
The above condition can be eliminated in view of a more general
formulation (see section 5). By taking the divergence of equation
(\ref{eq:varb}) we get for any $t$:
\begin{equation}\label{eq:divb}
{\rm div}{\bf B}~=~0
\end{equation}
provided the initial data are compatible with this constraint.
\par\smallskip

By defining  $\rho ={\rm div}{\bf E}$ and by taking the
divergence of equation (\ref{eq:vare}) it is straightforward to
get the continuity equation:
\begin{equation}\label{eq:continu}
\frac{\partial \rho}{\partial t}~=~-{\rm div}(\rho {\bf V})
\end{equation}
Equation (\ref{eq:continu}) is the first one of many other
conservation laws associated with the above system of equations.
This is quite an important prerogative, since, in the search of
solitonic solutions, all the possible quantities must be preserved
in the evolution. The important fact is that, even in vacuum
$\rho$ is allowed to be different from zero. Evidently, by
enforcing $\rho =0$, we come back with (\ref{eq:vare}) and
(\ref{eq:varb}) to the standard Maxwell's equations, but,
certainly, this is not the case we would like to discuss.  As a
matter of fact, it is known that solitary waves do not belong to
the space of solutions of Maxwell's equations (we come back on
this issue later in section 4). In the past, this led to the
conclusion that quantum phenomena cannot be modelled by the
classical equations of electromagnetism, so that, as we shall
mention in the coming section, numerous variants of the Maxwell's
system have been taken into account. The model we are describing
here differs from the others for its simplicity and for many
additional reasons we are going to detail.

\setcounter{equation}{0}
\section{Lagrangian analysis}

We start by introducing the potentials ${\bf A}$ and $\Phi$ such
that:
\begin{equation}\label{eq:potenz}
{\bf B}~=~{1\over c}~{\rm curl}{\bf A}~~~~~~~~~{\bf E}~=~-{1\over c}
{\partial {\bf A}\over \partial t}~-~\nabla \Phi
\end{equation}
In this way we guarantee equations (\ref{eq:vare})  and (\ref{eq:divb}).
Then, we consider the standard Lagrangian of electromagnetism:
\begin{equation}\label{eq:lagr}
{\cal L}~=~\Vert {\bf E}\Vert^2 -c^2\Vert {\bf B}\Vert^2
\end{equation}
It is known that Maxwell's equations are  linked to the stationary
points of the action function of ${\cal L}$. To check this, one
writes ${\cal L}$ in terms of $\Phi$ and ${\bf A}$ and then
differentiate with respect to the variations $\delta\Phi$ and
$\delta {\bf A}$, having compact support in space and time. A
typical way to carry out this proof (that we do not report here)
passes through the construction of the electromagnetic tensor
$F_{ik}$ (see for instance \cite{landau}). Thus, one discovers
that (\ref{eq:varb}) must be true together with the condition:
$~\rho={\rm div}{\bf E}=0$.

\par\smallskip

Unfortunately, imposing $\rho=0$ brings to a subspace of solutions
that do not include, for example, solitary waves with compact
support. Extensions are then necessary and the literature is rich
of results. An usual approach is based on a modification of the
Lagrangian, by adding a further term, on the basis of physical
considerations. With this respect, we mention the pioneering paper
of \cite{born} (see also \cite{yang} and \cite{yang2}). For a
more recent general viewpoint we refer for instance to
\cite{bencif2}, \cite{benci2} (see also \cite{badiale} for a
general review).
\par\smallskip

The procedure of adapting the Lagrangian is sometimes successful,
but the new corresponding system of equations turns out to be
heavily nonlinear and some of the good invariance properties of
Maxwell's equations might be lost. Theoretical results are usually
addressed to the existence of stable solitonic solutions. Another
approach is to couple some equations, directly derived from
Maxwell's system, with other type of equations, such as
Schr\"odinger or Klein-Gordon (see for example \cite{benci},
\cite{coclite}, \cite{dodd}).
\par\smallskip

The path followed here is to recover the set of model equations
(\ref{eq:vare})-(\ref{eq:varb})-(\ref{eq:varv}) from the
stationary points of the action function associated with the
standard Lagrangian in (\ref{eq:lagr}). In order to avoid the
condition $\rho=0$, we put a constraint to the variations. In this
way, since there are less degrees of freedom for the test
functions, we come out with a larger space of solutions.
\par\smallskip

To this end, let us impose the following relation:
\begin{equation}\label{eq:vvinc}
c{\bf A}~=~ \Phi {\bf V}
\end{equation}
Note that the above is not a gauge condition.  As a matter of fact
it will be somehow stronger (see later). From (\ref{eq:vvinc}),
due to (\ref{eq:normv}), we easily get the scalar constraint:
\begin{equation}\label{eq:vincsca}
c\hspace{.05truecm}\Phi~=~{\bf V}\cdot {\bf A}
\end{equation}
With the help of (\ref{eq:vvinc}), we can  actually obtain our set
of equations. The formal proof of this fact is given in
\cite{funaro}, using standard variational arguments, after
writing ${\cal L}$ in terms of the electromagnetic tensor
$F_{ik}$. Here we add some heuristic considerations.
\par\smallskip

By standard calculus one obtains:
$${\cal L}~=~\Vert {\bf E}\Vert^2 -c^2\Vert {\bf B}\Vert^2~=~
 {\bf E}\cdot\left( -{1\over c}{\partial {\bf A}\over
\partial t}-\nabla\Phi\right)-c~{\bf B}\cdot{\rm curl}{\bf A}$$
$$=-{1\over c}~{\partial \over \partial t}({\bf E}\cdot {\bf A})+
{1\over c}~{\partial {\bf E}\over \partial t}\cdot {\bf A}
- {\bf E}\cdot \nabla\Phi
-c~{\bf A}\cdot {\rm curl}{\bf B}-c~{\rm div}
({\bf A}\times {\bf B})$$
$$=~-{1\over c} ~{\partial\over \partial t}( {\bf E}\cdot {\bf A})~+~
{1\over c}\left( {\partial {\bf E}\over \partial t} ~-~c^2
{\rm curl}{\bf B}\right)\cdot {\bf A}~~~~~~~~~~~~~~~~$$
\begin{equation}\label{eq:catena}
~~~~~~~~~~~~-~ {\rm div}(\Phi {\bf E})~+~\Phi~ {\rm div}{\bf E}~
-~c~ {\rm div}({\bf A}\times {\bf B})
\end{equation}
At this point, by imposing condition (\ref{eq:vvinc}) and by noting that
${\bf E}\cdot {\bf A}=0$ (obtained by multiplying (\ref{eq:varv}) by ${\bf A}$),
the above equation becomes:
$${\cal L}~=~\Vert {\bf E}\Vert^2 -c^2\Vert {\bf B}\Vert^2
~=~- {\rm div}\Big( \Phi ({\bf E}~+~{\bf V}\times {\bf B})
\Big)$$
\begin{equation}\label{eq:catena2}
~+~{1\over c}\left( {\partial {\bf E}\over\partial t}~-
~c^2 {\rm curl}{\bf B}~+~ ({\rm div}{\bf E}){\bf V}\right)
\cdot {\bf A}
\end{equation}
We now observe that ${\cal L}$ vanishes when:
\begin{equation}\label{eq:sella}
\Vert {\bf E}\Vert ~=~c\Vert {\bf B}\Vert
\end{equation}
Consequently, if (\ref{eq:varv}) and (\ref{eq:vare}) hold true, we
are exactly in the situation ${\cal L}=0$, that is: all our waves
will have the norm of ${\bf E}$ equal to that of $c{\bf B}$. More
technical (but not too difficult) is to show that ${\cal L}=0$ is
actually a stationary point of the action function associated with
${\cal L}$.
\par\smallskip

Finally, we remark that the constraint  (\ref{eq:vincsca}) might
be included in the Lagrangian, as a penalty term, in the following
way:
\begin{equation}\label{eq:lagrp}
{\cal L}~=~\Vert {\bf E}\Vert^2 -c^2\Vert {\bf
B}\Vert^2~+~\frac{\rho}{c} ({\bf A} \cdot {\bf V} -
c\hspace{.05truecm}\Phi )
\end{equation}
The added term is similar to the one we would have in presence of
moving charged particles in an electromagnetic field (see, e.g.:
\cite{landau}).

\setcounter{equation}{0}
\section{Eikonal equation}

With a simple analysis, more general properties  of the solutions
can be recovered. For example, by expressing ${\bf E}$ and ${\bf
B}$ in term of the potentials, one can prove that:
\medskip
$$c^2({\bf E}+{\bf V}\times {\bf B})= -\frac{\partial}{\partial t}
(c{\bf A}-\Phi {\bf V})+{\bf V}~{\rm div}(c{\bf A}-\Phi {\bf V})~~~~~~
$$
\begin{equation}\label{eq:ebv}
~~~~~~~~~+{\bf V}\times {\rm curl}(c{\bf A}-\Phi {\bf V}) -c{\bf
V}\cdot\frac{D {\bf A}}{Dt}~+~\frac{\Phi}{2}~\nabla \Vert {\bf
V}\Vert^2
\end{equation}
where $D{\bf A} /Dt=\partial {\bf A}/\partial t+({\bf
V}\cdot\nabla ){\bf A}$ is the substantial derivative of ${\bf A}$
along the velocity vector field ${\bf V}$. Thanks to
(\ref{eq:vvinc}) and (\ref{eq:normv}), one concludes that:
\begin{equation}\label{eq:equiv}
{\bf E}~+~{\bf V}\times {\bf B}~=~0 ~~~~~\iff ~~~~~{\bf V}\cdot\frac{D{\bf A}}{Dt}~=~0
\end{equation}
Moreover, the following relations hold:
\begin{equation}\label{eq:relat}
{\bf E}\cdot {\bf V}=0 ~~~~~~~~~ {\bf E}\cdot {\bf B}=0 ~~~~~~~~~
 {\bf V}\cdot {\bf B}=0~~~~~~~~{\bf V}=c~\frac{{\bf E}\times {\bf B}}{\Vert{\bf E}\times {\bf B}\Vert}
\end{equation}
The first one is obtained by scalarly  multiplying (\ref{eq:varv})
by ${\bf V}$. The second one is similarly obtained by multiplying
(\ref{eq:varv}) by ${\bf B}$. The third and the fourth ones are
consequence of the two previous orthogonality relations together
with conditions  (\ref{eq:normv}) and (\ref{eq:sella}). Thus
$({\bf E}, {\bf B}, {\bf V})$ is a right-handed orthogonal
triplet.
\par\smallskip

Thanks to the above relations one can prove a classical results regarding
the time derivative of the energy:
\begin{equation}\label{eq:energ}
\frac{1}{2}\frac{\partial}{\partial t} \Big(\Vert {\bf E}\Vert^2
+c^2\Vert {\bf B}\Vert^2\Big)=
~c^2\big( {\rm curl} {\bf B}\cdot {\bf E}-{\rm curl} {\bf E}\cdot {\bf B}\big)
=- c^2~{\rm div}({\bf E}\times {\bf B})
\end{equation}
where ${\bf E}\times {\bf B}$ is the Poynting vector.
\par\smallskip

Additional conclusions are obtained by imposing  conditions on the
field ${\bf V}$. For example, one may require that ${\bf V}$ is
not subjected to transversal acceleration:
\begin{equation}\label{eq:subv}
~\frac{D {\bf V}}{Dt}~=~\frac{\partial {\bf V}}{\partial t}+({\bf
V}\cdot\nabla ){\bf V}~=~0
\end{equation}
A way to satisfy (\ref{eq:subv}) is to suppose that ${\bf V}$  is
an irrotational stationary field. This means that $~{\bf
V}=\nabla\Psi~$ for some scalar potential $\Psi$. In this way:
$D{\bf V} /Dt=\nabla ( \partial\Psi /\partial t +\frac{1}{2}\nabla
\Vert\Psi\Vert^2)=0$, yielding (\ref{eq:subv}). With this choice,
the most important consequence is the following eikonal equation:
\begin{equation}\label{eq:eiko2}
\Vert\nabla\Psi\Vert~=~c
\end{equation}
directly obtainable from (\ref{eq:normv}). This ensures that our
solutions develop according to the rules of geometrical optics.
Note also that, if we use (\ref{eq:vvinc}) and (\ref{eq:subv}), one gets: $cD{\bf A}/Dt =
D (\Phi{\bf V})/Dt={\bf V}(D \Phi/Dt)$. Therefore, the condition at
the right-hand side of (\ref{eq:equiv}) can be replaced by:
\begin{equation}\label{eq:dphi}
\frac{D\Phi}{Dt}~=~\frac{\partial\Phi}{\partial t}~+~{\bf V}\cdot\nabla \Phi~=~0
\end{equation}
In conclusion, when ${\bf V}$ is irrotational and $~D\Phi /Dt=0$,
we automatically have (\ref{eq:varv}) and (\ref{eq:eiko2}),
showing that the wave-fronts evolve as prescribed by geometrical
optics.
\par\smallskip

If, in addition to these hypotheses, one also requires the following Lorenz
gauge conditions on the potentials:
\begin{equation}\label{eq:lorenz}
{\rm div}{\bf A}~+~{1\over c}{\partial \Phi\over \partial t}~=~0
\end{equation}
from (\ref{eq:vvinc}), we also get another continuity equation:
\begin{equation}\label{eq:contfi}
{\partial \Phi\over\partial t}~=~-{\rm div}(\Phi {\bf V})
\end{equation}
which is stronger than demanding $~D \Phi/Dt =0$.
\par\smallskip

Our final purpose is to show that it is possible to explicitly
compute infinite interesting solutions of the system
(\ref{eq:vare})-(\ref{eq:varb})-(\ref{eq:varv}), realizing an
extended range of conservation laws.

\setcounter{equation}{0}
\section{Explicit solutions}

\noindent In the cartesian reference frame $(x,y,z)$, let us take
the following solution candidates:
$${\bf E}=\Big(E_1(x,y)g(z-ct),~ E_2(x,y)g(z-ct),~ 0\Big)$$
\begin{equation}\label{eq:solc}
{\bf B}=\Big(B_1(x,y)g(z-ct),~ B_2(x,y)g(z-ct),~ 0\Big)
\end{equation}
representing two field distributions, modulated by the  function
$g$, laying on parallel planes and shifting  in the direction of
the $z$-axis at the speed of light.  The functions $E_1$, $E_2$,
$B_1$ and $B_2$ are smooth on the whole plane $(x,y)$ (they may
allowed, for example, to be zero outside a domain of finite
measure). The function $g$ is also smooth.
\par\smallskip

By setting $~{\bf V}=(0,0,c)~$ and by direct substitution into
the equations (\ref{eq:vare})-(\ref{eq:varb})-(\ref{eq:varv}), the
vector fields shown in (\ref{eq:solc}) are solutions under the
following assumptions:
\begin{equation}\label{eq:ass1}
E_1=cB_2~~~~~~~~~~E_2=-cB_1~~~~~~~~~{\rm div}{\bf B}=\partial B_1/\partial x +
\partial B_2/\partial y=0
\end{equation}
for any choice of $g$. With this setting, one can actually find a function $A_3$
and construct the two potentials:
\begin{equation}\label{eq:pot1}
\Phi=A_3(x,y)g(z-ct) ~~~~~~~  {\bf A}=\Big(0,~0,~A_3(x,y)g(z-ct)\Big)
\end{equation}
It is straightforward to check that all the conditions
(\ref{eq:vvinc}), (\ref{eq:sella}), (\ref{eq:relat}),
(\ref{eq:energ}), (\ref{eq:subv}), (\ref{eq:eiko2}),
(\ref{eq:dphi}), (\ref{eq:lorenz}), (\ref{eq:contfi}) are
verified. Thus, depending on the arbitrary functions $A_3$ and
$g$,  we can build infinite solutions of our system. If $A_3$ and
$g$ have compact support, we get solitary electromagnetic waves
shifting unperturbed at the speed of light. Note that $\rho ={\rm
div}{\bf E}\not =0$ and (\ref{eq:continu}) becomes the trivial
transport equation:
\begin{equation}\label{eq:transp}
\frac{\partial \rho}{\partial t}~=~-c~\frac{\partial
\rho}{\partial z}
\end{equation}
\par\smallskip

If we instead start from the general setting in  (\ref{eq:solc}),
and try to solve the full set of Maxwell's equations (i.e.,
including the additional condition ${\rm div}{\bf E}=0$), there
are no chances of getting interesting nontrivial solutions. In
this case we can forget about the velocity vector ${\bf V}$. By
direct substitution, one easily deduces that both the functions
$~E_1-iE_2~$ and $~B_1-iB_2$, where $i$ is the imaginary unit,
must be holomorphic (entire) on the whole complex plane $x+iy$
(see, for instance \cite{conway}). By the Liouville's theorem:
{\sl if an entire holomorphic function is bounded, then it is a
constant}, we entail that there are no bounded continuous
electromagnetic fields of the form (\ref{eq:solc}), having finite
energy and solving the whole set of Maxwell's equations, with the
exception of $~{\bf E}=0$ and ${\bf B}=0$ (note that classical
plane waves have not finite energy). We recall that solutions may
exist if we assume that ${\bf E}$ and ${\bf B}$ do not belong to
the tangent plane of the advancing front. In this situation,
however, the Poynting vector ${\bf E}\times{\bf B}$, indicating
the direction of the energy flow (see (\ref{eq:energ})), is not
lined up with the direction of movement, in contrast with the
rules of geometrical optics. This brings to a diffusive behavior.
\par\smallskip

We can now transform the  equations
(\ref{eq:vare})-(\ref{eq:varb})-(\ref{eq:varv}) in spherical
coordinates $(r, \theta ,\phi )$ and take the following fields,
distributed on the tangent planes of spherical wave-fronts:
$${\bf E}={1\over r}\Big(0,~ E_2(\theta ,\phi)g(r-ct),~ E_3(\theta ,\phi )g(r-ct)\Big)$$
\begin{equation}\label{eq:solc2}
{\bf B}={1\over r}\Big(0,~B_2(\theta ,\phi )g(r-ct),~
B_3(\theta ,\phi )g(r-ct)\Big)
\end{equation}
With this choice, the energy  density $~\Vert {\bf E}\Vert^2
+c^2\Vert {\bf B}\Vert ^2~$ remains constant when integrated over
any spherical surface. Similarly to the previous case, by defining
${\bf V}=(c,0,0)$, we can get infinite solutions, provided they
satisfy:
\begin{equation}\label{eq:ass2}
E_3=-cB_2~~~~~~~~E_2=cB_3~~~~~~~~{\rm div}{\bf B}=
\partial (B_2\sin \theta )/\partial\theta +\partial B_3 /\partial\phi
\end{equation}
allowing us to build the potentials:
\begin{equation}\label{eq:pot2}
\Phi = A_1(\theta ,\phi)g(r-ct)~~~~~~{\bf A}=\Big( A_1(\theta ,\phi)g(r-ct), ~0, ~0\Big)
\end{equation}
where $A_1$ and $g$ are arbitrary. As before all the conditions
(\ref{eq:vvinc}), (\ref{eq:sella}), (\ref{eq:relat}),
(\ref{eq:energ}), (\ref{eq:subv}), (\ref{eq:eiko2}),
(\ref{eq:dphi}), (\ref{eq:lorenz}), (\ref{eq:contfi}) are
verified. By taking $A_1(\theta ,\phi)=-\cos\theta$ and
$g(s)=\sin\omega s$, among the solutions we recognize the one,
corresponding to the monochromatic field generated by an
infinitesimal dipole, usually employed in applications (see
\cite{bleaney}, \cite{bornw}, \cite{joos}):
\begin{equation}\label{eq:solc3}
{\bf E}=\Big(0,~ \frac{\sin\theta}{r} \sin \omega (r-ct),~
0\Big)~~~~~~ {\bf B}=\Big(0,~0,~\frac{\sin\theta}{r} \sin \omega
(r-ct)\Big)
\end{equation}

Searching instead for solutions of the Maxwell's  system, one
finds out that the complex functions $~E_2-iE_3\sin \theta~$ and
$~B_2-iB_3\sin \theta~$ should be holomorphic on the Riemann
sphere. Hence, ${\bf E}$ and ${\bf B}$ are bounded if and only if
they are zero. Therefore, the only possible solutions of Maxwell's
equations in vacuum, having bounded electromagnetic fields laying
on tangent planes of spherical fronts are identically zero. As a
consequence the one in (\ref{eq:solc3}) cannot be solution (in
fact ${\rm div}{\bf E}\not =0$). The same arguments can be applied
to any closed, bounded, compact, oriented surface. Non trivial
solutions, for example the celebrated Hertz solutions (see
\cite{hertz}, \cite{bornw}, \cite{joos}), are instead possible
by allowing the fields to have  radial components $E_1$ and $B_1$.
In this way, however, the Poynting vector ${\bf E}\times {\bf B}$
is not aligned with the direction of movement and the evolution of
the corresponding wave-fronts do not comply with the rules of
geometrical optics (i.e.: the surfaces obtained as the envelope of
the electromagnetic vectors, representing the wave-fronts, are not
spherical), so that many of the conservation laws reported here
are not satisfied.

\setcounter{equation}{0}
\section{Concluding remarks}

With the new model equations we have an extended range of
solutions, not available in the Maxwellian case. In particular,
this includes wave-packets of almost any form, both from the
viewpoint of the shape of the wave-fronts and the information
written on them. These wave-packets travel unperturbed at speed
$c$, along the direction of the vector ${\bf V}$, having constant
norm. If ${\bf V}$ is stationary and irrotational, then the
eikonal equation is verified and the wave-fronts perfectly follow
the laws of geometrical optics. The equations are compatible with
the stationary points of the standard Lagrangian of the
electromagnetism. In this new framework, the differentiation has
to be taken with respect to potentials subject to a certain
constraint, that, thanks (\ref{eq:ebv}), is equivalent in practice
to impose (\ref{eq:varv}).
\par\smallskip

At this point, something has to be said about  the physical
implications. First of all, we recall that in \cite{funaro}
equation (\ref{eq:varv}) is generalized as follows:
\begin{equation}\label{eq:varv2}
{D{\bf V}\over Dt}~=~-\mu\big({\bf E}~+~{\bf V}\times{\bf
B}\big)~-~{\nabla p \over\rho}
\end{equation}
which is the Euler equation for compressible fluids (recall the
continuity equation (\ref{eq:continu})), with an electromagnetic
type forcing term. The scalar $p$ is a suitable pressure and $\mu$
is a  constant whose dimension is {\sl charge/mass}. This approach
combines the evolution of electromagnetic entities with that of a
(non material) inviscid fluid. Some numerical experiments,
regarding the interaction of waves with matter (i.e., the
diffraction due to the passage of a photon through a small hole)
are examined in \cite{funaro2}. Other numerical experiments,
concerning electromagnetic waves trapped in bounded regions of
space (basically, forming vortex rings) are reported in
\cite{chinosi}. In this new context, that is better suited for a
general relativity framework, we must drop condition
(\ref{eq:normv}) and replace it with an eikonal equation in a
suitable metric space. In order to maintain the exposition at
basic level, we do not add further considerations concerning this
generalization.
\par\smallskip

So far, we studied the special case when $p=0$ and $D{\bf
V}/Dt=0$. In this circumstance, the fluid moves unperturbed. From
the electromagnetic point of view, equation (\ref{eq:varv}) says
that the ``Lorentz force'' acting on the wave is null. This is in
agreement with the fact that our  solitons freely travel, without
the influence of external factors.
\par\smallskip

Equation (\ref{eq:vare}) is the Amp\`ere law,  without explicit
external currents, but with a sort of  electric density developing
together with the wave and satisfying automatically the continuity
equation (\ref{eq:continu}). The presence of a nonvanishing $\rho$
is not in contrast with the Gauss's divergence theorem. If our
travelling soliton has a bounded support included in a solid
region $\Omega$, then the integral $~\int_{\delta\Omega} {\bf
E}\cdot {\bf n}=\int_\Omega \rho~$ is equal to zero. Therefore,
seen as a whole, our soliton is not a charge (it does not even
emit electromagnetic fields during its movement), although,
inside, there are points where $\rho\not =0$.
\par\smallskip

In order to deal with equation (\ref{eq:varv2}) we may generalize the
Lagrangian in (\ref{eq:lagrp}) by setting:
\begin{equation}\label{eq:lagrpg}
{\cal L}~=~\Vert {\bf E}\Vert^2 -c^2\Vert {\bf
B}\Vert^2~+~\frac{\rho}{c} ({\bf A} \cdot {\bf V} -
c\hspace{.05truecm}\Phi )~+~\frac{c\rho}{\mu}\sqrt{c^2-\Vert {\bf
V}\Vert^2} ~-~\frac{p}{\mu}
\end{equation}
We come back to the special case of equations (\ref{eq:lagrp}) by
imposing $p=0$ and recalling (\ref{eq:normv}). This approach takes
also into account relativistic effects. We did not carry out any
analysis concerning the above Lagrangian. Perhaps, the techniques
introduced in \cite{badiale} to study the general Lagrangian
${\cal L}_{u,S,{\bf A},\Phi}$ may also be applied in our
situation.

\par\smallskip

To conclude we say that the most important achievement, realized
with the new  model equations, is to be able to produce,
maintaining a physical meaning,  vector waves that display all the
peculiarities of standard electromagnetic emissions. In addition,
waves with the characteristics of a photon are allowed by the
model, providing the link between the evolution of electromagnetic
entities and that of classical mechanical bodies. Finally, let us
point out that the new system of equations can be proved to be
invariant under Lorentz transformations (see \cite{funaro},
section 2.6), can be written in covariant form (see
\cite{funaro}, chapter 4) and is naturally linked to the
derivative of the electromagnetic stress tensor (see
\cite{funaro}, section 4.2, and \cite{donev}).

\end{document}